\documentclass[12pt]{article}
\usepackage{pstricks,pst-node,pstcol}
\def\sqr#1#2{{\vcenter{\hrule height.#2pt
      \hbox{\vrule width.#2pt height#1pt \kern#1pt
          \vrule width.#2pt}
      \hrule height.#2pt}}}
\def\twosqr#1#2{{\vcenter{\hrule height.#2pt
      \hbox{\vrule width.#2pt height#1pt \kern#1pt
          \vrule width.#2pt}
      \hrule height.#2pt
      \hbox{\vrule width.#2pt height#1pt \kern#1pt
          \vrule width.#2pt}
      \hrule height.#2pt}}}
\def\thrsqr#1#2{{\vcenter{\hrule height.#2pt
      \hbox{\vrule width.#2pt height#1pt \kern#1pt
          \vrule width.#2pt}
      \hrule height.#2pt
\hbox{\vrule width.#2pt height#1pt \kern#1pt
          \vrule width.#2pt}
      \hrule height.#2pt
      \hbox{\vrule width.#2pt height#1pt \kern#1pt
          \vrule width.#2pt}
      \hrule height.#2pt}}}

\definecolor{dgray}{gray}{0.8}
\definecolor{sgray}{gray}{0.7}
\definecolor{tgray}{gray}{0.5}
\definecolor{fgray}{gray}{0.3}
\definecolor{figray}{gray}{0.1}

\begin{document}
\newfont{\elevenmib}{cmmib10 scaled\magstep1}%
\newcommand{\tabtopsp}[1]{\vbox{\vbox to#1{}\vbox to12pt{}}}
\font\larl=cmr10 at 24pt
%\font\got=eufb10 at 12pt
\newcommand{\es}{\got s}

\newcommand{\preprint}{
            \begin{flushleft}
   \elevenmib Yukawa\, Institute\, Kyoto\\
            \end{flushleft}\vspace{-1.3cm}
            \begin{flushright}\normalsize  \sf
            YITP-01-42\\
           {\tt hep-th/0105197} \\ May 2001
            \end{flushright}}
\newcommand{\Title}[1]{{\baselineskip=26pt \begin{center}
            \Large   \bf #1 \\ \ \\ \end{center}}}
\hspace*{2.13cm}%
\hspace*{0.7cm}%
\newcommand{\Author}{\begin{center}\large \bf
           V.\,I.\, Inozemtsev\footnote{
permanent address: BLTP JINR, 141980 Dubna, Moscow Region, Russia}
 and R.\, Sasaki \end{center}}
\newcommand{\Address}{\begin{center}
            Yukawa Institute for Theoretical Physics\\
     Kyoto University, Kyoto 606-8502, Japan
      \end{center}}
\newcommand{\Accepted}[1]{\begin{center}{\large \sf #1}\\
            \vspace{1mm}{\small \sf Accepted for Publication}
            \end{center}}
\baselineskip=20pt

\preprint
\thispagestyle{empty}
\bigskip
\bigskip
\bigskip

\Title{Hierarchies of Spin Models related to Calogero-Moser Models}
\Author

\Address
\vspace{1cm}

\begin{abstract}
The universal formulation of  spin exchange models related to
Calogero-Moser
models implies the existence of integrable hierarchies, which
have not been explored. We show the general structures and features of the
spin exchange model hierarchies by taking as examples
the well-known Heisenberg
spin chain with the nearest neighbour interactions.
The energy spectra of the second
member  of the hierarchy belonging to the models based on the $A_r$ root
systems $(r=3,4,5)$ are  explicitly and {\em exactly\/} evaluated. They
show many many interesting features and in particular,
 much higher degree of degeneracy than the original Heisenberg  model,
 as expected from the integrability.
\end{abstract}
\bigskip
\bigskip
\bigskip

%%%%%%%%%%%%%%%

In a previous paper \cite{is1} we
 presented universal Lax pairs for the spin Calogero-Moser
 models \cite{suthsha2,fmp,HikWa} and spin exchange  models
\cite{halsha,ino1,simal,ino2} which give rise to enough
number of conserved quantities. As in the continuous Calogero-Moser models
\cite{Cal,Sut,CalMo}, these conserved quantities are shared by
all members of the
integrable models belonging to the same {\em hierarchy}.
However, in the context of
spin exchange models which we will mainly discuss in this paper, the notion
of
integrable hierarchies is rather unfamiliar and virtually unexplored in
contradistinction with the soliton hierarchies or Calogero-Moser
hierarchies.
The purpose of the present paper is to call attention to this interesting
and
potentially  useful subject in mathematical/statistical
physics by examining the
second member of the integrable hierarchy of the simple
and well-known Heisenberg
spin chain with the nearest neighbour interactions.
The energy eigenvalues of these models are evaluated explicitly and
exactly  for
relatively small  numbers of the spins and remarkably high degree of {\em
degeneracies} of the eigenvalues is observed. We believe that this fact
is  closely related to the existence of a large number of
conserved quantities shared
by the hierarchy.

The hierarchy of the spin exchange models \cite{is1} consists of models
associated with a fixed root system $\Delta$ and the totality of irreducible
representations %$\{{\cal R}\}$'s
of the finite reflection (or Coxeter or Weyl) group
$G_\Delta$.
In the case of the Heisenberg spin chain the root system is
$A_r$, which corresponds
to the Lie algebra of $su(r+1)$. In this case the Weyl group of $A_r$ is the
well-known symmetric group ${\cal S}_{r+1}$.
First let us rephrase  various concepts and dynamical
quantities of the well-known
Heisenberg spin chain in terms of the languages of the
weights and roots and the
associated reflections, which are useful for the universal description of
the
Calogero-Moser models \cite{OP1,DHoker_Phong,bcs2,bms} and the spin exchange
models \cite{is1}.
If the
$A_r$ root system is embedded in
${\bf R}^{r+1}$, it has the following simple representation in terms of an
orthonormal   basis $\{{\bf e}_j\}$ of ${\bf R}^{r+1}$:
\begin{equation}
A_r=\{{\bf e}_j-{\bf e}_k|\  j,k=1,\ldots,r+1\}.
\end{equation}
The original Heisenberg model is based on the
{\em vector} (${\bf V}$) representation of $A_r$
corresponding to the simplest Young
diagram
$\sqr{9}5$. It consists of the following {\em weights\/}:
\begin{equation}
\sqr{9}5:\qquad {\bf V}=\{{\bf e}_j+\gamma|\  j=1,\ldots,r+1\},
\label{vecwei}
\end{equation}
in which $\gamma$ is orthogonal to all the $A_r$ roots and it will be
omitted
hereafter.
At each  ``site" ${\bf e}_j$ (or simply the site $j$)
a spin space is attached.
Throughout this paper we consider the simplest case of
the $1/2$ spins of $su(2)$,
that is ups ($\uparrow$)  and downs ($\downarrow$).
In other words, the state space of the original
Heisenberg chain is a $2^{r+1}$
dimensional vector space.
The Hamiltonian now reads
\begin{equation}
{\cal H}=J\sum_{\rho\in\Pi^{(1)}}\hat{\cal P}_\rho,\quad J\in{\bf R},
\label{hamdef}
\end{equation}
in which $J$ is the coupling constant and $\Pi^{(1)}$
is the set of {\em affine
simple roots\/} of $A_r$:
\begin{equation}
\Pi^{(1)}=\{{\bf e}_1-{\bf e}_2,{\bf e}_2-{\bf e}_3,\ldots,
{\bf e}_r-{\bf e}_{r+1},
{\bf e}_{r+1}-{\bf e}_1\},
\label{afroots}
\end{equation}
and $\hat{\cal P}_\rho$ is the spin exchange operator
corresponding to the root
$\rho$.
In the present case $\hat{\cal P}_{{\bf e}_j-{\bf e}_{j+1}}$
exchanges spins at the
sites $j$ and $j+1$ (to be more precise the sites ${\bf e}_j$ and
${\bf e}_{j+1}$)
and keeps the other spins intact. Let us denote the Pauli spin
matrix acting on the
site ${\bf e}_j$ simply by $\vec{\sigma}_j$, then  $\hat{\cal P}_\rho$ has a
well-known realisation:
\begin{equation}
\rho={\bf e}_j-{\bf e}_k\Longrightarrow
\hat{\cal P}_\rho=(1+\vec{\sigma}_j\cdot\vec{\sigma}_k)/2.
\label{Pdef1}
\end{equation}
Thus we find that the above Hamiltonian  (\ref{hamdef}) describes the
nearest
neighbour Heisenberg spin chain and the last root in (\ref{afroots})
provides
a periodic lattice.

It is worth remarking that the above Hamiltonian (\ref{hamdef})
can be obtained as
a special limit of a more general spin exchange model with elliptic exchange
functions \cite{ino1}, and the conserved quantities can be constructed
\cite{ino2}
as shown by one of the authors. The above Hamiltonian (\ref{hamdef}) and its
conserved quantities are shared by all the members of the same integrable
hierarchy.
Each member of the hierarchy,  to be called a model ${\cal R}$ for short, is
specified by an irreducible representation ${\cal R}$ of the Weyl group of
$A_r$.
The elements of ${\cal R}$, to be called ``sites", too, are vectors in
${\bf R}^{r+1}$ which form a single Weyl orbit:
\begin{equation}
\mu\in{\cal R}\Longrightarrow s_{\rho}(\mu)\in{\cal R},\quad
 \forall\rho\in\Delta,
\end{equation}
in which $s_\rho$ reflects
a vector $\xi\in{\bf R}^{r+1}$ in a hyperplane perpendicular to the root
$\rho$:
\begin{equation}
\xi\to s_\rho(\xi)=\xi-(\rho^\vee\!\cdot\xi)\rho,\quad
\rho^\vee\equiv 2\rho/\rho^2,\quad \rho\in\Delta.
\end{equation}
As above, an $su(2)$ spin, an up ($\uparrow$)
or a down ($\downarrow$) is attached
to each ``site" $\mu$. Thus the state space of the
model ${\cal R}$ is a $2^{D}$
dimensional vector space in which $D$ is the total
number of sites in ${\cal R}$.
The action of the spin exchange operator
$\hat{\cal P}_\rho$ is now
\begin{equation}
\hat{\cal P}_\rho:\quad \mbox{spin at}\ \mu\Leftrightarrow
\mbox{spin at}\ s_\rho(\mu),\quad \forall\mu\in{\cal R}
\end{equation}
simultaneously. If a site $\mu$ is orthogonal to a given root $\rho$,
$\rho\cdot\mu=0$, then $\mu=s_\rho(\mu)$.
That is, $\hat{\cal P}_\rho$ affects only those spins at
the sites $\mu$ such that
$\rho\cdot\mu\neq0$.
In this way the action of the Hamiltonian (\ref{hamdef}) in
the model
${\cal R}$ is completely defined.

The simplest example of the hierarchy is given by the
{\em anti-symmetric tensor\/}
representation (${\bf AT}$) of $A_r$ corresponding to the Young diagram
$\twosqr{9}5$\,. It consists of $D=r(r+1)/2$ weights of the form
\begin{equation}
\twosqr{9}5:\qquad {\bf AT}=\{{\bf e}_j+{\bf e}_k|\  k>j=1,\ldots,r+1\},
\label{tenwei}
\end{equation}
in which as before (\ref{vecwei}) we have omitted a constant
vector $\delta$ which is
orthogonal to all the roots.
For notational simplicity, we express the site ${\bf e}_j+{\bf e}_k$ by a
symmetric pair of integers $(j,k)=(k,j)$ and the Pauli spin matrix
on the site by
$\vec{\sigma}_{(j,k)}$.
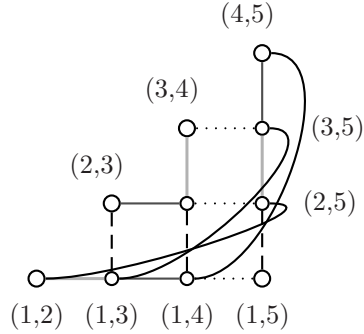
\begin{figure}[tbp]
    \centering
    % remove from here to the next \includegraphics{spin.eps}
\begin{pspicture}(0,0)(5.5,5)
\cnode(1.5,0.5){0.1}{A}
\cnode(2.5,0.5){0.1}{B}
\cnode(2.5,1.5){0.1}{C}
\cnode(3.5,1.5){0.1}{D}
\cnode(3.5,2.5){0.1}{E}
\cnode(0.5,0.5){0.12}{F}
\cnode(1.5,1.5){0.12}{G}
\cnode(2.5,2.5){0.12}{H}
\cnode(3.5,3.5){0.12}{I}
\cnode(3.5,0.5){0.12}{J}
\rput(1.5,0){\footnotesize (1,3)}
\rput(2.5,0){\footnotesize (1,4)}
\rput(3.5,0){\footnotesize (1,5)}
\rput(0.5,0){\footnotesize (1,2)}
\rput(1.3,2.0){\footnotesize (2,3)}
\rput(2.3,3.0){\footnotesize (3,4)}
\rput(3.3,4.0){\footnotesize (4,5)}
\rput(4.4,1.5){\footnotesize (2,5)}
\rput(4.5,2.5){\footnotesize (3,5)}
\ncline[linestyle=dashed]{A}{G}
\ncline[linestyle=dashed]{B}{C}
\ncline[linestyle=dashed]{J}{D}
%the above e_{1}-e_{2} 0.9
\ncline[linewidth=0.4mm,linecolor=sgray]{F}{A}
\ncline[linewidth=0.4mm,linecolor=sgray]{C}{H}
\ncline[linewidth=0.4mm,linecolor=sgray]{D}{E}
%the above e_{2}-e_{3} 0.7
\ncline[linecolor=fgray]{A}{B}
\ncline[linecolor=fgray]{G}{C}
\ncline[linecolor=fgray]{E}{I}
%the above e_{3}-e_{4} 0.5
\ncline[linestyle=dotted]{B}{J}
\ncline[linestyle=dotted]{C}{D}
\ncline[linestyle=dotted]{H}{E}
\nccurve{F}{D}
\nccurve{A}{E}
\nccurve{B}{I}
% the above e_{5}-e_{1} 0.1
\end{pspicture}
    \caption{Structure of the second member of the $A_{4}^{(1)}$
Heisenberg spin model hierarchy with nearest neighbour interactions.
The spins on the sites which are connected by the same bonds are
exchanged simultaneously.
}
    \label{fig:1}
\end{figure}
Obviously the sites $\{(j,k)\}$'s form a two dimensional orthogonal lattice.
A generic site $(j,k)$ is connected to {\em four\/} nearest neighbouring
sites:
\begin{equation}
(j,k) \Longleftrightarrow (j-1,k),\ (j+1,k),\ (j,k-1),\ (j,k+1).
\end{equation}
Those sites on the ``boundary"
\begin{equation}
\mbox{boundary of}\ {\twosqr{9}5}=
\{(1,2),\ldots,(j,j+1),\ldots,(r,r+1),(r+1,1)\}
\end{equation}
are connected to {\em two\/} other sites.
The spin exchange operator $\hat{\cal P}_\rho$ for
$\rho={\bf e}_j-{\bf e}_k$ exchanges the spins in
the same {\em row\/} or {\em
column\/}, i.e.
$(j,\ell)\Leftrightarrow (k,\ell)$, $\forall\ell\neq j,k$, {\em
simultaneously\/}:
\begin{equation}
\rho={\bf e}_j-{\bf e}_k,\quad \hat{\cal P}_\rho:\qquad
{\bf e}_j+{\bf e}_\ell\Leftrightarrow
{\bf e}_k+{\bf e}_\ell,\quad \ell\neq j,k.
\end{equation}
Thus for the {\em anti-symmetric tensor\/} (${\bf AT}$) model, the
same Hamiltonian (\ref{hamdef}) no longer describes the nearest neighbour
interactions in the ordinary sense. As in the original Heisenberg model
(\ref{Pdef1})  the spin exchange operator has a
simple realisation in terms of the
Pauli spin matrices:
\begin{equation}
\rho={\bf e}_j-{\bf e}_k\Longrightarrow
\hat{\cal P}_\rho=\prod_{\ell\neq
j,k}[(1+\vec{\sigma}_{(j,\ell)}\cdot\vec{\sigma}_{(k,\ell)})/2].
\label{Pdef2}
\end{equation}
It should be emphasised that the two realisations
(\ref{Pdef1}) and (\ref{Pdef2}) of
the  spin exchange operator $\hat{\cal P}_\rho$ (and any other realisations)
satisfy the same commutation relations
\begin{equation}
  \hat{\cal P}_{\rho}\hat{\cal P}_{\sigma}\hat{\cal P}_{\rho}
   =\hat{\cal P}_{s_{\rho}(\sigma)},\quad
\hat{\cal P}_{\rho}^2=1,\quad \hat{\cal P}_{-\rho}=\hat{\cal
P}_{\rho},
\label{pcom}
\end{equation}
which are a direct consequence of the well-known commutation relations of
the
reflections
\begin{equation}
{s}_{\rho}{s}_{\sigma}{s}_{\rho}
   ={s}_{s_{\rho}(\sigma)},\quad
{s}_{\rho}^2=1,\quad {s}_{-\rho}={s}_{\rho}.
\end{equation}

Since the
Hamiltonian commutes with the total spin operator
\begin{equation}
\vec{S}={1\over2}\sum_{\mu\in{\cal R}}\vec{\sigma}_\mu,
\end{equation}
the operators $S^2$ and $S_z$ define good quantum numbers.
In the absence of the
external magnetic field, the energy eigenstates
consist of $su(2)$ multiplets of
spin $S$ ($2S+1$ fold degenerate) states  as in
the original Heisenberg spin chain.
  The vector (${\bf V}$),
anti-symmetric tensor (${\bf AT}$)
and the other higher rank anti-symmetric tensor representations
of $A_r$ are special
examples of the {\em minimal\/} representations which have played
important roles in
the Calogero-Moser Lax pairs \cite{DHoker_Phong,bcs1,bcs2,bms,cfs1}.
The Calogero-Moser Lax pairs in these representations were first presented
by
D'Hoker and Phong \cite{DHoker_Phong}.
Minimal representations are characterised by
the condition (see \cite{bcs1,bcs2})
\begin{equation}
\rho^\vee\!\cdot\mu=0,\ \pm1,\quad \forall \mu\in{\cal R},\quad
\forall\rho\in\Delta.
\end{equation}
It is expected that the spin exchange models for these ${\cal R}$'s have
some
simpler features than others.

In the rest of this paper we show the energy eigenvalues of the
anti-symmetric
tensor representation (${\bf AT}$) model for lower rank ($r\le5$) cases.
We evaluate the Hamiltonian (\ref{hamdef}) in each $S_z=k$, $-r/2\le k\le
r/2$
sector. In each sector, the Hamiltonian is a non-negative symmetric matrix
with integer entries. In both ${\bf V}$ and
${\bf AT}$ models or any other models
in the hierarchy, the highest eigenvalue is $r+1$ which is the number of
spin
exchange operators
$\hat{\cal P}_\rho$ in  the Hamiltonian (\ref{hamdef}). The lowest possible
eigenvalue is
$-(r+1)$. These two facts are based on the relation
$\hat{\cal P}_\rho^2=1$ (\ref{pcom}) implying that its eigenvalues are
$\pm1$.
The highest eigenvalue $r+1$ occurs in every $S_z$
sector but the lowest possible
eigenvalue $-(r+1)$ appears only in smaller $S_z$
sector which have larger number of states. In particular, in the original
Heisenberg (${\bf V}$) model,
the lowest possible eigenvalue $-(r+1)$ never occurs,
see (\ref{a3v}), (\ref{a4v}), (\ref{a5v}) and (\ref{a9v}).
Once the lowest possible eigenvalue $-(r+1)$ is reached in an $S_z=k$
sector,  then in all the $|S_z|<k$ sectors no new eigenvalues will appear.
The energy eigenvalues of the vector representation
(${\bf V}$) models are also
listed for comparison.
(The coupling constant $J$ is put to unity for simplicity.)
It is easy to see that the energy eigenvalues of the
{\em single spin down\/}
({\em up\/}) sector of the (${\bf V}$) model are given by
\begin{equation}
{\sqr{9}5}\ (\downarrow):\quad {\cal E}_j=
r-1+2\cos{2j\pi\over{(r+1)}},\quad j=0,1,\ldots,r,
\end{equation}
which are related to the eigenvalues of the Cartan matrix of $A_r^{(1)}$ and
thus related to the mass spectrum of the affine Toda field theory based on
$A_r^{(1)}$ \cite{atft}.
All these eigenvalues appear in other sectors because of
the $su(2)$ invariance. It is also easy to see that for proper choices of
the bases
the Hamiltonian of the
{\em double spin down\/} ({\em up\/})
sector of the (${\bf V}$) model is exactly the same as that of
{\em single spin down\/} ({\em up\/})
sector of the (${\bf AT}$) model and so on:
\begin{equation}
{\sqr{9}5} \ (\downarrow\downarrow)\ ={\twosqr{9}5} \ (\downarrow),\quad
{\sqr{9}5} \ (\downarrow\downarrow\downarrow)\
={\thrsqr{9}5} \ (\downarrow),\quad
\cdots, .
\end{equation}

\paragraph{\underline{$A_2$ case}} In this case the vector rep.  $\sqr{9}5$
(${\bf V}$)  model and the
anti-symmetric tensor rep. $\twosqr{9}5$
(${\bf AT}$) model are both 3-dimensional
and these two models are equivalent from the spin model point of view.
The total number of states is $8=2^3$ and
the energy eigenvalues  together with
({\tt multiplicity}) are
\begin{equation}
3\ ({\tt 4}),\quad 0\ ({\tt 4}).
\end{equation}
The former consists of one $S=3/2$ multiplet,
whereas the zero energy states consist of two $S=1/2$ states.

\paragraph{\underline{$A_3$ case}} The vector rep.
$\sqr{9}5$ is 4-dimensional
whereas the anti-symmetric tensor rep. $\twosqr{9}5$ is 6-dimensional.
The spectrum of the $\sqr{9}5$ (${\bf V}$) model is
\begin{equation}
\sqr{9}5:\quad 2^4=16\ \mbox{states}\quad
4\ ({\tt 5}), \quad 2\ ({\tt 7}), \quad 0\ ({\tt 3}), \quad -2\ ({\tt 1}).
\label{a3v}
\end{equation}
The maximal eigenvalue ${\cal E}=4$ states are one
$S=2$ multiplet and the ${\cal
E}=2$ states consist of two $S=1$ multiplets and one $S=0$ singlet.
The ${\cal E}=0$ states are one $S=1$ multiplet and
the lowest energy ${\cal E}=-2$
state is a singlet. The spectrum of the $\twosqr{9}5$ (${\bf AT}$) model is
\begin{equation}
\twosqr{9}5:\quad 2^6=64\ \mbox{states}\quad
4\ ({\tt 11}), \quad 2\ ({\tt 26}), \quad 0\ ({\tt 12}), \quad -2\ ({\tt
14}),
\quad -4\ ({\tt 1}).
\end{equation}
The maximal eigenvalue ${\cal E}=4$ states are one $S=3$,
$S=1$ multiplet each and
a singlet.
The ${\cal E}=2$ states are three $S=2$ and three $S=1$ multiplets and two
singlets.
The zero energy  ${\cal E}=0$ states are one $S=2$ multiplet,
two $S=1$ multiplets
and one singlet.
The ${\cal E}=-2$ states have one $S=2$ and three $S=1$ multiplets. The
lowest
energy ${\cal E}=-4$ state is a singlet.

\paragraph{\underline{$A_4$ case}} The vector rep.
$\sqr{9}5$ is 5-dimensional
whereas the anti-symmetric tensor rep. $\twosqr{9}5$ is 10-dimensional.
The spectrum of the $\sqr{9}5$ (${\bf V}$) model is
\begin{equation}
\sqr{9}5:\quad 2^5=32\ \mbox{states}\quad 5\ ({\tt 6}),
\quad {5\pm\sqrt{5}\over2}\
({\tt 8}),\quad 1\pm\sqrt{5}\ ({\tt 4}),\quad 1\ ({\tt 2}).
\label{a4v}
\end{equation}
The maximal eigenvalue ${\cal E}=5$ states are one $S=5/2$ multiplet.
The pair of energy levels ${\cal E}=(5\pm\sqrt{5})/2$ are both two $S=3/2$
multiplets.
Another pair ${\cal E}=1\pm\sqrt{5}$ of states, one of which is the
lowest energy,
have  each two $S=1/2$ multiplets. The ${\cal E}=1$ states are one $S=1/2$
multiplet. The spectrum of the $\twosqr{9}5$  (${\bf AT}$) model is
\begin{eqnarray}
\twosqr{9}5:\quad 2^{10}&=&1024\ \mbox{states}\quad 5\ ({\tt 34}),\quad
{5\pm\sqrt{5}\over2}\ ({\tt 112}),\quad 1\pm\sqrt{5}\ ({\tt 116}),\quad
-1\pm\sqrt{5}\ ({\tt 76}),\nonumber\\
&& {-5\pm\sqrt{5}\over2}\ ({\tt 32}),\quad \pm\sqrt{5}\ ({\tt 36}),
\quad 1\ ({\tt
58}),\quad 0\ ({\tt 144}),\quad -1\ ({\tt 38}),\quad -5\ ({\tt 6}).
\end{eqnarray}
This is a highly degenerate spectrum.
The average multiplicity per energy level is
$1024/15\approx68$ in contrast to
$1024/139\approx 7.3$ in the vector rep. of $A_9$ case
(\ref{a9vec}) which has the
same number of sites as in this case. The
highest
${\cal E}=5$ states consist of {\tt 1} $S=5$ (11-dimensional) , {\tt 1}
$S=3$
(7-dimensional) , {\tt 2}
$S=2$ (5-dimensional)  and {\tt 2}
$S=1$ (3-dimensional) multiplets, which we denote as follows:
\begin{equation}
{\cal E}=5\quad ({\tt 34}):\quad {\tt 1} \times {\bf 11},
\qquad {\tt 1} \times {\bf
7},\quad\ {\tt 2}\times {\bf 5},\quad {\tt 2}\times {\bf 3}.
\end{equation}
Similarly we have the following degeneracy pattern:
\begin{equation}
\begin{array}{llrrrrrr}
{\cal E}\ \ =&{(5\pm\sqrt{5})/2}&({\tt 112}):&{\tt 2} \times {\bf 9},&
{\tt 4} \times {\bf 7},& {\tt 8} \times {\bf 5},& {\tt 8} \times {\bf
3},& {\tt 2} \times {\bf 1},\\[4pt]
{\cal E}\ \ =&{1\pm\sqrt{5}}& ({\tt 116}):& {\tt 2} \times
{\bf 9},& {\tt 4} \times {\bf 7},& {\tt 8} \times {\bf 5},&
{\tt 8} \times {\bf 3},& {\tt 6} \times {\bf 1},\\[4pt]
{\cal E}\ \ =&{-1\pm\sqrt{5}}& ({\tt 76}):&  &{\tt 4}
\times {\bf 7},& {\tt 4} \times {\bf 5},& {\tt 8} \times {\bf
3},& {\tt 4} \times {\bf 1},\\[4pt]
{\cal E}\ \ =&{(-5\pm\sqrt{5})/2}& ({\tt 32}):& &
& {\tt 4} \times {\bf 5},&
{\tt 4}
\times {\bf 3},& \\[4pt]
{\cal E}\ \ =&{\pm\sqrt{5}}& ({\tt 36}):& & {\tt 1} \times
{\bf 7},&{\tt 3} \times {\bf 5},& {\tt 4} \times {\bf 3},&
{\tt 2} \times {\bf 1},\\[4pt]
{\cal E}\ \ =&1 & ({\tt
58}):&\ {\tt 1}
\times {\bf 9},& {\tt 2} \times {\bf 7},& {\tt 4} \times {\bf
5},& {\tt 4} \times {\bf 3},& {\tt 3} \times {\bf 1},\\[4pt]
{\cal E}\ \ =&0& ({\tt 144}):& &  {\tt 4} \times {\bf
7},& {\tt 12} \times {\bf 5},& {\tt 16}
\times {\bf 3},& {\tt 8} \times {\bf 1},\\[4pt]
{\cal E}\ \ =&-1& ({\tt 38}):& &  {\tt 2} \times {\bf
7},& {\tt 2} \times {\bf 5},& {\tt 4}
\times {\bf 3},& {\tt 2} \times {\bf 1},\\[4pt]
{\cal E}\ \ =&{-5}& ({\tt 6}):& && {\tt 1}
\times {\bf 5},&
&{\tt 1} \times {\bf 1}.\nonumber
\end{array}
\end{equation}
In the $A_4$ models all the multiplicities are even numbers.

\paragraph{\underline{$A_5$ case}} The vector rep.
$\sqr{9}5$ is 6-dimensional
whereas the anti-symmetric tensor rep. $\twosqr{9}5$ is 15-dimensional.
The spectrum of the $\sqr{9}5$ (${\bf V}$) model is
\begin{eqnarray}
{\sqr{9}5}:\quad 2^6=64\ \mbox{states}\ &&
6\ ({\tt 7}={\tt 1}\times{\bf 7}),
\quad 5\ ({\tt 10}={\tt2}\times{\bf5}), \quad 4\
({\tt 3}={\tt1}\times{\bf3}),
\quad 3\ ({\tt 10}={\tt2}\times{\bf5}),\nonumber\\
&&\ 2\ ({\tt 7}={\tt2}\times{\bf3}+{\tt1}\times{\bf1}), \quad 1\ ({\tt
6}={\tt2}\times{\bf3}),
\quad 0\ ({\tt 1}),
\label{a5v}\\
&&\ (5\pm\sqrt{17})/2\ ({\tt 6}={\tt2}\times{\bf3}),
\quad 1\pm\sqrt{5}\ ({\tt
3}={\tt1}\times{\bf3}),
\quad 1\pm\sqrt{13}\ ({\tt 1}).\nonumber
\end{eqnarray}
The spectrum of the $\twosqr{9}5$ (${\bf AT}$) model is
\begin{eqnarray}
&&{\twosqr{9}5}: 2^{15}=32768\ \mbox{states}\ \quad
6\ ({\tt 156}), \quad 5\ ({\tt 776}), \quad 4\
({\tt 1060}),
\quad 3\ ({\tt 776}),
\ 2\ ({\tt 2756}), \quad 1\ ({\tt 4224})\nonumber\\
&& \quad 0\ ({\tt 1304}),
\ -1\ ({\tt 3600}), \quad -2\ ({\tt
1988}),
\quad -3\ ({\tt  296}),\quad -4\ ({\tt 628}),
\quad -5\ ({\tt 296}),\quad -6\ ({\tt 28}),
\nonumber\\
&&\ (5\pm\sqrt{17})/2\ ({\tt 1200}),\quad \ (-5\pm\sqrt{17})/2\
({\tt 576}), \quad 1\pm\sqrt{5}\ ({\tt
600}),\quad -1\pm\sqrt{5}\ ({\tt 288}),
\nonumber\\
&&\quad 1\pm\sqrt{13}\ ({\tt 264}),
\quad -1\pm\sqrt{13}\ ({\tt 240}), \quad \pm c_1,
\pm c_2, \pm c_3\ ({\tt 1424}),
\end{eqnarray}
in which
\begin{equation}
c_j=\left(1-10\cos[(\phi-2j\pi)/3]\right)/3,\quad j=1,2,3,
\quad \cos\phi={7/{250}}.
\end{equation}
These six eigenvalues always appear with the same
multiplicities in each $S_z$ sector.
This is a highly degenerate spectrum.
The degeneracy pattern is as follows:
\begin{equation}
\begin{array}{crrrrrrrrr}
{\cal E}&{\tt total\ Mult.}&{\bf 16}&{\bf14}&{\bf12}&{\bf10}&{\bf8}
&{\bf 6}&{\bf 4}&{\bf2}\\
6   & {\tt 156  }       &       1       &               &       1       &
3       &       4       &       6       &       6       &       3       \\
5       &       {\tt    776     }       &               &       2
&       4       &       10      &       24      &       34      &
36      &       30      \\
4       &       {\tt    1060    }       &               &       1
&       4       &       13      &       30      &       50      &
61      &       42      \\
3       &       {\tt    776     }       &               &       2
&       4       &       10      &       24      &       34      &
36      &       30      \\
2       &       {\tt    2756    }       &               &       1
&       8       &       33      &       76      &       135     &
168     &       113     \\
1       &       {\tt    4224    }       &               &       2
&       12      &       46      &       118     &       210     &
256     &       182     \\
0       &       {\tt1304        }       &               &
&       3       &       15      &       34      &       66      &
85      &       55      \\
-1      &       {\tt    3600    }       &               &
&       6       &       34      &       100     &       186     &
234     &       168     \\
-2      &       {\tt    1988    }       &               &
&       2       &       19      &       52      &       104     &
136     &       95      \\
-3      &       {\tt    296     }       &               &
&               &       2       &       8       &       16      &
20      &       18      \\
-4      &       {\tt    628     }       &               &
&               &       5       &       17      &       33      &
45      &       32      \\
-5      &       {\tt    296     }       &               &
&               &       2       &       8       &       16      &
20      &       18      \\
-6      &       {\tt    28      }       &               &               &
&               &       1       &       2       &       1       &       2
\\
 (5\pm\sqrt{17})/2      &       {\tt    1200    }       &               &
 2       &       6       &       16      &       34      &       54      &
 64      &       44      \\
 (-5\pm\sqrt{17})/2     &       {\tt    576     }       &               &
 &               &       4       &       16      &       30      &       42
 &       30      \\
 1\pm\sqrt{5}   &       {\tt    600     }       &               &       1
 &       3       &       8       &       17      &       27      &       32
 &       22      \\
 -1\pm\sqrt{5}  &       {\tt    288     }       &               &
 &               &       2       &       8       &       15      &       21
 &       15      \\
 1\pm\sqrt{13}  &       {\tt    264     }       &               &
 &       1       &       4       &       6       &       13      &       17
 &       9       \\
-1 \pm\sqrt{13} &       {\tt    240     }       &               &
&       1       &       3       &       6       &       12      &       15
&       9       \\
 \pm c_1,\ \pm c_2,\ \pm c_3    &       {\tt    1424    }       &
 &               &       4       &       14      &       40      &       74
 &       86      &       64.
\end{array}
\end{equation}

The average multiplicity per energy level is $32768/31\approx1057$.
We believe this is a rule for the higher members of the spin model
hierarchies.

\paragraph{\underline{$A_9$ case}} The vector rep.
$\sqr{9}5$ is 10-dimensional
which has the same number of sites as the anti-symmetric
tensor rep. $\twosqr{9}5$
of $A_4$. There are 139 different energy levels among 1024 states.
Here we list the
spectrum (numerical values only) and ({\tt multiplicity}) for comparison.
\begin{eqnarray}
&& {\sqr{9}5}:\quad 2^{10}=1024\ \mbox{states} \nonumber\\
&&10     \ ({\tt 11 }),\
9.618  \ ({\tt 18 }),\
9.345  \ ({\tt 14 }),\
9.173  \ ({\tt 10 }),\
9.118  \ ({\tt 1 }),\
9.099  \ ({\tt 6 }),\
9.064  \ ({\tt 7 }),\
8.982  \ ({\tt 3 }),\nonumber\\
&&8.800  \ ({\tt 10 }),\
8.678  \ ({\tt 14 }),\
8.666  \ ({\tt 5 }),\
8.618  \ ({\tt 18 }),\
8.613  \ ({\tt 10 }),\
8.250  \ ({\tt 6 }),\
8.191  \ ({\tt 14 }),\nonumber\\
&&8  \ ({\tt 7 }),\
7.978  \ ({\tt 3 }),\
7.963  \ ({\tt 2 }),\
7.934  \ ({\tt 14 }),\
7.855  \ ({\tt 10 }),\
7.800  \ ({\tt 6 }),\
7.594  \ ({\tt 6 }),\
7.421  \ ({\tt 2 }),\nonumber\\
&&7.382  \ ({\tt 18 }),\
7.361  \ ({\tt 10 }),\
7.346  \ ({\tt 10 }),\
7.303  \ ({\tt 10 }),\
7.206  \ ({\tt 10 }),\
7.169  \ ({\tt 6 }),\
7.120  \ ({\tt 14 }),\nonumber\\
&&6.893  \ ({\tt 6 }),\
6.730  \ ({\tt 6 }),\
6.695  \ ({\tt 7 }),\
6.677  \ ({\tt 1 }),\
6.653  \ ({\tt 14 }),\
6.540  \ ({\tt 2 }),\
6.501  \ ({\tt 10 }),\nonumber\\
&&6.382  \ ({\tt 18 }),\
6.347  \ ({\tt 6 }),\
6.341  \ ({\tt 3 }),\
6.212  \ ({\tt 1 }),\
6.141  \ ({\tt 6 }),\
6.131  \ ({\tt 6 }),\
6.112  \ ({\tt 10 }),\
6.061  \ ({\tt 10 }),\nonumber\\
&&6  \ ({\tt 26 }),\
5.902  \ ({\tt 10 }),\
5.899  \ ({\tt 10 }),\
5.708  \ ({\tt 14 }),\
5.627  \ ({\tt 10 }),\
5.618  \ ({\tt 6 }),\
5.459  \ ({\tt 14 }),\
5.364  \ ({\tt 2 }),\nonumber\\
&&5.357  \ ({\tt 14 }),\
5.213  \ ({\tt 10 }),\
5.198  \ ({\tt 6 }),\
5.167  \ ({\tt 10 }),\
5.145  \ ({\tt 10 }),\
5.132  \ ({\tt 6 }),\
5.107  \ ({\tt 6 }),\
5.093  \ ({\tt 2 }),\nonumber\\
&&5.018  \ ({\tt 10 }),\
4.929  \ ({\tt 6 }),\
4.902  \ ({\tt 1 }),\
4.848  \ ({\tt 14 }),\
4.666  \ ({\tt 10 }),\
4.648  \ ({\tt 3 }),\
4.646  \ ({\tt 2 }),\
4.618  \ ({\tt 6 }),\nonumber\\
&&4.599  \ ({\tt 5 }),\
4.591  \ ({\tt 6 }),\
4.561  \ ({\tt 10 }),\
4.514  \ ({\tt 10 }),\
4.421  \ ({\tt 14 }),\
4.390  \ ({\tt 10 }),\
4.356  \ ({\tt 6 }),\
4.341    \ ({\tt 2 }),\nonumber\\
&&4.232  \  ({\tt 14 }),\ 4.171
\  ({\tt 6 }),\
4.163  \  ({\tt 6 }),\
4  \  ({\tt 12 }),\
3.939  \  ({\tt 6 }),\
3.697  \  ({\tt 10 }),\
3.566  \  ({\tt 10 }),\
3.563  \  ({\tt 2 }),\nonumber\\
&&3.488  \  ({\tt 2 }),\
3.431  \  ({\tt 2 }),\
3.382  \  ({\tt 6 }),\
3.284  \  ({\tt 10 }),\
3.275  \  ({\tt 6 }),\
3.202  \  ({\tt 6 }),\
3.199  \  ({\tt 10 }),\
3.189  \  ({\tt 6 }),\nonumber\\
&&3.091  \  ({\tt 14 }),\
3.027  \  ({\tt 6 }),\
3.025  \  ({\tt 1 }),\
2.964  \  ({\tt 14 }),\
2.869  \  ({\tt 1 }),\
2.747  \  ({\tt 6 }),\
2.694  \  ({\tt 6 }),\
2.687  \  ({\tt 3 }),\nonumber\\
&&2.517  \  ({\tt 10 }),\
2.471  \  ({\tt 10 }),\
2.382  \  ({\tt 6 }),\
2.241  \  ({\tt 7 }),\
2.208  \  ({\tt 10 }),\
2.137  \  ({\tt 10 }),\
2.096  \  ({\tt 3 }),\
1.923  \  ({\tt 2 }),\nonumber\\
&&1.916  \  ({\tt 2 }),\
1.904  \  ({\tt 10 }),\
1.696  \  ({\tt 6 }),\
1.687  \  ({\tt 5 }),\
1.646  \  ({\tt 2 }),\
1.565  \  ({\tt 6 }),\
1.523  \  ({\tt 6 }),\
1.454  \  ({\tt 10 }),\nonumber\\
&&1.140  \  ({\tt 6 }),\
1.106  \  ({\tt 6 }),\
1.078  \  ({\tt 6 }),\
0.962  \  ({\tt 1 }),\
0.942  \  ({\tt 6 }),\
0.808  \  ({\tt 1 }),\
0.744  \  ({\tt 10 }),\
0.359  \  ({\tt 10 }),\nonumber\\
&&0.332  \  ({\tt 3 }),\
0.122  \  ({\tt 3 }),\
0.025  \  ({\tt 10 }),\
-0.029  \  ({\tt 2 }),\
-0.535  \  ({\tt 2 }),\
-0.561  \  ({\tt 6 }),\
-0.773  \  ({\tt 2 }),\nonumber\\
&&-0.782  \  ({\tt 6 }),\
-0.952  \  ({\tt 5 }),\
-1.492  \  ({\tt 6 }),\
-2.087  \  ({\tt 6 }),\
-2.541  \  ({\tt 1 }),\
-3.184  \  ({\tt 3 }),\
-4.031  \  ({\tt 1 }).\nonumber\\
\label{a9vec}
\label{a9v}
\end{eqnarray}

\bigskip
Some remarks and comments are in order.

One might feel that the
{\em simultaneous exchange}  of the spins is artificial.
Rather it is  quite
natural within the context of the conventional spin exchange models, if we
consider any member of the conserved quantities as a new
Hamiltonian but keeping
the sites of the spin the same as in the original model.
As shown in the previous paper \cite{is1} the conserved quantities consist
of the linear combination of ordered products of the spin exchange
operators.
Each product term of the conserved quantity specifies
simultaneous exchange of
spins. The only exception is the original
spin exchange models based on $A_r$ root
systems and the vector representation ${\bf V}$.
For the other root systems, even
for the classical ones $B_r$, $C_r$ or $D_r$,
the single spin exchange operator
implies a multiple exchange as shown explicitly in (3.21) of \cite{is1}.

It would be interesting to determine the energy eigenvalues of the
Hamiltonian
(\ref{hamdef}) for a different choice of the set
of affine simple roots, for example
$D_r^{(1)}$:
\begin{equation}
\Pi^{(1)}=\{{\bf e}_1-{\bf e}_2,{\bf e}_2-{\bf e}_3,\ldots,
{\bf e}_{r-1}-{\bf e}_{r}, {\bf e}_{r-1}+{\bf
e}_{r},-({\bf e}_{1}+{\bf e}_2)\},
\label{drafroots}
\end{equation}
and for the $D_r$ vector representations or (anti-) spinor representations.
The former would give a $D$-series analog of
the nearest neighbour Heisenberg model.

Another well-known method for spin models is the use of Yang-Baxter
equations
\cite{kulskl}. It would be interesting to compare the fusion procedure for
the
$R$-matrices with the present method of generating hierarchies.

\paragraph{Acknowledgements}
V.\,I.\,I. is supported by JSPS long term fellowship.
R. S. is partially
supported  by the Grant-in-aid from the Ministry of Education, Culture,
Sports, Science and Technology, Japan,  priority area (\#707)
``Supersymmetry and unified theory of elementary particles".
%%%%%%%%%%%%%%%%%%%%%%%%%%%%%%%%%%%%%%%%%%%%%%%%%%%%%


\begin{thebibliography}{99}
\bibitem{is1}
V.\,I.\, Inozemtsev and R.\, Sasaki,
``Universal Lax pairs for spin Calogero-Moser models and spin
exchange models", Kyoto preprint YITP-01-34, {\tt hep-th/0105164}, May 2001.


\bibitem{suthsha2}
B.\,S.\, Shastry and B.\, Sutherland, ``Superlax pairs and infinite
symmetries
in the $1/r^2$ system", Phys. Rev. Lett. {\bf 70} (1993) 4029-4033;
B.\, Sutherland and B.\,S.\, Shastry, ``Solutions of some integrable
one-dimensional quantum system",
Phys. Rev. Lett. {\bf 71} (1993) 5-8.

\bibitem{fmp}
A.\,P.\,Polychronakos, ``Exchange operator formalism for integrable systems
of
particles", Phys. Rev. Lett. {\bf 69} (1992) 703-705;
M.\,Fowler and J.\,A.\,Minahan, ``Invariants of the Haldane-Shastry $SU(N)$
chain", Phys. Rev. Lett. {\bf 70} (1993) 2325-2328;
A.\,P.\,Polychronakos,
``Lattice integrable systems of Haldane-Shastry type",
{\it ibid} {\bf 70} (1993)
2329-2331.



\bibitem{HikWa}
K.\, Hikami and   M.\, Wadati,  ``Integrability of Calogero-Moser
spin system", J. Phys. Soc. Jpn.
 {\bf 62} (1993) 469-472.

\bibitem{halsha}
F.\, D.\, M.\,  Haldane, ``Exact Jastrow-Gutzwiller resonating
    valence bond ground state of the spin 1/2 antiferromagnetic
    Heisenberg chain with 1/r**2 exchange",
Phys. Rev. Lett. {\bf 60} (1988) 635-638; B.\,S.\,
Shastry, ``Exact solution of $S=1/2$ Heisenberg antiferromagnetic chain
with long-ranged interactions", {\it ibid} {\bf 60} (1988) 639-642.


\bibitem{ino1}
V.\,I.\,Inozemtsev, ``On the connection between the one-dimensional
$S=1/2$ Heisenberg chain and Haldane-Shastry model",
J. Stat. Phys. {\bf 59} (1990) 1143-1156.




\bibitem{simal}
B.\,D.\,Simons and B.\,L.\,Altschuler,
``Exact ground state of an open $S={1/2}$ long-range Heisenberg
antiferromagnetic spin chain",
Phys. Rev. {\bf B50} (1994) 1102-1105;
D.\,Bernard, V.\,Pasquier and D.\,Serban, ``Exact solution of
 long-range interacting spin chains with boundaries",
Europhys. Lett. {\bf 30} (1995) 301-306;
T.\,Yamamoto, ``Multicomponent Calogero model of $B_N$-type confined in
a harmonic potential", Phys. Lett. {\bf A208} (1995) 293;
T.\,Yamamoto and O.\,Tsuchiya, ``Integrable $1/r^2$ spin chain with
reflecting end", J. Phys. {\bf A29} (1996) 3977-3984, {\tt
cond-mat/9602105}.

\bibitem{ino2}
V.\,I.\,Inozemtsev, ``Invariants of linear combinations of transpositions",
Lett. Math. Phys. {\bf 36} (1996) 55-63.

\bibitem{Cal}  F.\, Calogero, ``Solution of the one-dimensional
\(N\)-body problem with quadratic and/or inversely quadratic pair
potentials", J. Math. Phys. {\bf 12} (1971) 419-436.
\bibitem{Sut}
B.\, Sutherland, ``Exact results for a quantum many-body problem in
one-dimension. II'', Phys. Rev. {\bf A5} (1972) 1372-1376.
\bibitem{CalMo}
J.\, Moser, ``Three integrable Hamiltonian systems connected with
isospectral deformations'',  Adv. Math. {\bf 16} (1975) 197-220;\
J.\, Moser,  ``Integrable systems of non-linear evolution equations",
in {\it Dynamical Systems, Theory and Applications\/};\
J. Moser, ed., Lecture Notes in Physics {\bf 38} (1975),
Springer-Verlag;\
F.\,Calogero, C.\, Marchioro and O.\, Ragnisco, ``Exact solution of the
classical and quantal one-dimensional many body problems with
the two body potential \(V_{a}(x)=g^2a^2/\sinh^2\,ax\)'', Lett. Nuovo
Cim. {\bf 13} (1975) 383-387;\
F.\,Calogero,``Exactly solvable one-dimensional many body problems'',
Lett. Nuovo Cim. {\bf 13} (1975) 411-416.

%\bibitem{spinmodels}

\bibitem{OP1} M.\,A.\, Olshanetsky and A.\,M.\, Perelomov,
``Completely integrable Hamiltonian systems connected with
 semisimple Lie algebras",
 Inventions Math. {\bf 37} (1976), 93-108;
 ``Classical integrable finite-dimensional systems related to Lie
 algebras'',
 Phys. Rep.  {\bf C71} (1981), 314-400.

 \bibitem{DHoker_Phong}
E.\,D'Hoker and D.\,H.\,Phong, ``Calogero-Moser
Lax pairs with spectral parameter for general Lie algebras'',
Nucl. Phys. {\bf B530} (1998) 537-610, {\tt hep-th/9804124}.
%%CITATION = HEP-TH 9804124;%%




\bibitem{bcs1}
 A.\,J.\, Bordner, E.\, Corrigan and R.\, Sasaki,
``Calogero-Moser models I: a new formulation'',
Prog. Theor. Phys. {\bf 100} (1998) 1107-1129, {\tt hep-th/9805106};


 \bibitem{bcs2}  A.\,J.\, Bordner, E.\, Corrigan and R.\, Sasaki,
``Generalized Calogero-Moser models and  universal Lax pair operators'',
 Prog. Theor. Phys. {\bf 102}  (1999)  499-529,
 {\tt  hep-th/9905011}.
%%CITATION = HEP-TH 9905011;%%



\bibitem{bms}  A.\,J.\, Bordner, N.\,S.\, Manton and R.\, Sasaki,
``Calogero-Moser models V:  Supersymmetry,
and Quantum Lax Pair", Prog. Theor. Phys. {\bf 103} (2000) 463-487,
{\tt hep-th/9910033}.
%%CITATION = HEP-TH 9910033;%%


\bibitem{cfs1}
R.\,Caseiro, J.-P.\,Fran\c{c}oise and R.\,Sasaki,
``Algebraic Linearization of Dynamics of Calogero Type for any
Coxeter Group'',
J.\ Math.\ Phys.\ {\bf 41} (2000) 4679-4986,
{\tt hep-th/0001074}.
%%CITATION = HEP-TH 0001074;%%



\bibitem{atft}
A.\, E.\, Arinshtein, V.\, A.\, Fateev, A.\,B.\,Zamolodchikov,
``Quantum S-matrix of the 1+1 dimensional Toda chain",
Phys. Lett. {\bf B87} (1979) 389-392;
H.\, W.\, Braden, E.\, Corrigan, P.\,E.\,Dorey\ and R.\, Sasaki,
``Affine Toda field theory and exact S-matrices",
Nucl. Phys. {\bf B338} (1990) 689-746.




\bibitem{kulskl}
P.\, P.\,Kulish and E.\, K.\,Sklyanin,  ``Quantum spectral transform method.
Recent developments",  Lecture Notes in Phys., {\bf 151}, Springer,
Berlin-New York, (1982)  61-119.


\end{thebibliography}
\end{document}